\documentclass[a4paper,twoside,twocolumn,english,prl,aps,showpacs,floatfix,groupedaddress,superscriptaddress]{revtex4}
\usepackage{mathptmx}
\usepackage[]{fontenc}
\usepackage[latin9]{inputenc}
\usepackage{array}
\usepackage{verbatim}
\usepackage{longtable}
\usepackage{graphicx}
\usepackage{amssymb}

\makeatletter

\providecommand{\tabularnewline}{\\}

\usepackage{babel}
\makeatother

\begin{document}

\title{Pfaffian State Generation by Strong 3-Body Dissipation }

\author{M. Roncaglia}

\affiliation{Max-Planck-Institut f\"ur Quantenoptik, Hans-Kopfermann-Str. 1,
D-85748, Garching, Germany}

\author{M. Rizzi}

\affiliation{Max-Planck-Institut f\"ur Quantenoptik, Hans-Kopfermann-Str. 1,
D-85748, Garching, Germany}

\author{J.I. Cirac}

\affiliation{Max-Planck-Institut f\"ur Quantenoptik, Hans-Kopfermann-Str. 1,
D-85748, Garching, Germany}

\begin{abstract}
We propose a scheme for preparing and stabilizing the Pfaffian state
with high fidelity in rapidly rotating 2D traps containing a small
number of bosons. The goal is achieved by strongly increasing 3-body
loss processes, which suppress superpositions of three particles while
permitting pairing. This filtering mechanism gives rise to reasonably
small losses if the system is initialized with the right angular momentum.
We discuss some methods for tuning 3-body interactions independently
of 2-body collisions. 
\end{abstract}


\maketitle
Trapped cold atoms represent a new frontier for the study of many
body phases, due to higher versatility if compared to ordinary condensed
matter systems \citep{Bloch08}. Confining an atomic gas to live in
a 2D geometry represents a benchmark for testing quantum Hall effects
(QHE), where fast rotation for bosons plays formally the same role
as the magnetic field for electrons confined in eterojunctions \citep{Fetter08}.
Strictly speaking, the analogy with the QHE is attained when the centrifugal
force equals the radial trapping force, recovering the big degeneracy
of the lowest Landau level (LLL). Nonetheless, some phases can be
stabilized keeping the rotation below this limit \citep{wilkin00}. 

One of the most intriguing states appearing in the QHE is the Pfaffian,
first introduced by Moore and Read \citep{MooreRead91} in the context
of paired Hall states. Successively, it was proposed as a candidate
for describing the filling factor $\nu=5/2$ in fermionc QHE, although
the question is still under debate and the Pfaffian state has not
been observed so far. However, this special wavefunction captures
the attention of the scientific community, since it has the peculiar
property of non-Abelian braiding statistics of its anyonic excitations.
This feature discriminates the Pfaffian from other QHE states (like
the Laughlin) and makes it attractive in the context of topological
quantum computation (TQC) \citep{Kitaev03,DasSarmaRevModPhys08}. 

The aim of this Letter is to propose and investigate a novel method
for preparing and stabilizing the Pfaffian state in rotating harmonic
traps loaded with cold bosonic atoms. Our proposal is based on exploiting
dissipation instead of suffering from it \citep{syassen_science08,Garcia-Ripoll09,Daley09},
by implementing a filtering procedure that basically projects a state
with the right angular momentum onto the desired Pfaffian. In experiments,
this can be achieved by a readily feasible mechanism to increment
the relative importance of 3-body losses with respect to 2-body elastic
scattering: namely, squeezing the trap \citep{Schneider_science08}.
Furthermore, perfect projection is attained by turning off 2-body
collisions (by Feshbach resonances \citep{roati_nature08}). The method
of preparing the initial state with a given angular momentum is not
crucial for the filtering to work, but we expect it to be possible
in the near future\emph{ }by putting few atoms in single wells of
an optical lattice into rotation \citep{popp04}, or by realizing
gauge potentials \citep{Gunter09}.

The paper is organized as follows. First, we briefly revise the QHE
regime requirements for cold bosons and the physics of the LLL, which
is dominated by the nature of interactions. Hereby we demonstrate
the different behavior of 2- and 3-body contact potential under density
rescaling in 2D. The ground state (GS) diagram in the case of conservative
3-body scattering is presented, with particular attention to the Pfaffian-like
sector. This motivates the need for an effective 3-body repulsion.
The dissipative projection mechanism is then proven without 2-body
collisions, and investigated numerically in their presence. Finally
we suggest some experimental signatures and a method to prepare the
initial state.

At low energy, a realization of an effective 2D system is obtained
by setting the longitudinal trap frequency much bigger than the transverse
one, $\omega_{\perp}\gg\omega$, in order to freeze longitudinal motion
in the GS. The filling factor is defined as $\nu=N/l_{\mathrm{max}}$
with $l_{\mathrm{max}}$ the maximum angular momentum occupied by
single particles. In the frame rotating at angular speed $\Omega\hat{z}$,
the single body Hamiltonian in the trap can be written as \[
\mathcal{H}_{\mathrm{trap}}=\frac{\left(\vec{p}-\vec{A}\right)^{2}}{2m}+\frac{m}{2}(\omega^{2}-\Omega^{2})(x^{2}+y^{2}),\]
with $\vec{A}=m\Omega\hat{z}\times\vec{r}$. In the limit of centrifugal
deconfinement $\Omega\to\omega$, only the Coriolis force is remaining
and the system is formally equivalent to bosons of charge $q$ in
uniform magnetic field $\vec{B}=(2m\Omega/q)\hat{z}$. The one body
eigenfunctions in the LLL take a simple form when written in terms
of the complex coordinate $z=(x+iy)/\xi$, with $\xi=\sqrt{\hbar/m\omega}$:

\[
\psi_{n}(z)=\frac{1}{\sqrt{\pi n!}}z^{n}e^{-\left|z\right|^{2}/2},\]
and have energies $E_{n}=\hbar n(\omega-\Omega)\equiv l_{n}\:\delta\omega$
where $l_{n}$ are the angular momenta projections. Other Landau levels
(LL) are separated by a gap~$\sim2\omega$, and the LLL restriction
is claimed to be even more valid for higher-$L$ QHE states, due to
their stronger correlations \citep{morris06}. All the numerical calculations
presented below have been checked by including the first LL and verifying
that occupation is small there.

The physics in the LLL is dominated by the nature and the strength
of interactions, which drive the system into different filling factors
\citep{Fetter08}. In the context of cold bosonic gases in 2D, 2-particles
interactions can be modeled by contact potentials\begin{equation}
\mathcal{H}_{\mathrm{2}}=g_{2}^{\mathrm{2D}}\sum_{i<j}\delta^{(2)}(\vec{x}_{i}-\vec{x}_{j}),\label{eq:2-body delta potential}\end{equation}
with $g_{2}^{\mathrm{2D}}=\sqrt{8\pi}\hbar\omega\xi^{2}a/\xi_{\perp}$,
being $a$ the $s$-wave scattering length in 3D and $\xi_{\perp}=\sqrt{\hbar/m\omega_{\perp}}$
the longitudinal trap size \citep{Bloch08}. Analogously, pointlike
3-body interactions can be written in the following way \citep{Huang87,Greiter91}
\begin{equation}
\mathcal{H}_{\mathrm{3}}=g_{3}^{\mathrm{2D}}\sum_{i<j<k}\delta^{(2)}(\vec{x}_{i}-\vec{x}_{j})\delta^{(2)}(\vec{x}_{j}-\vec{x}_{k})\,,\label{eq:3-body delta potential}\end{equation}
neglecting for the moment their physical origin. Let us call K2 and
K3 the kernel of 2-body Eq.(\ref{eq:2-body delta potential}) and
3-body Eq.(\ref{eq:3-body delta potential}) term, respectively. Inside
K2, the Laughlin state with $\nu=1/2$ \begin{equation}
\Psi_{\mathrm{Lau}}=\prod_{i<j}(z_{i}-z_{j})^{2},\label{eq:laughlin}\end{equation}
has the lowest total angular momentum $L_{\mathrm{Lau}}=N(N-1)$,
and thus energy $E_{\mathrm{Lau}}=L_{\mathrm{Lau}}\delta\omega$.
{[}As usual, in (\ref{eq:laughlin}) and in the subsequent QHE wavefunctions
we omit the ubiquitous exponential and normalization factors]. Of
course, K2$\subset$K3 and (\ref{eq:laughlin}) is also annihilated
by the 3-body interaction, despite not being the GS. The lowest $L$
state in K3 is indeed the Pfaffian \citep{MooreRead91} \begin{equation}
\Psi_{\mathrm{Pf}}=\mathrm{Pf}\left(\frac{1}{z_{i}-z_{j}}\right)\prod_{i<j}(z_{i}-z_{j}),\label{eq:moore-read}\end{equation}
with $L_{\mathrm{Pf}}=N(N-2)/2$ and $\nu=1$. The prefactor $\mathrm{Pf}\left(\frac{1}{z_{i}-z_{j}}\right)$
makes possible the superposition of pairs, and it is formally equivalent
to a projected $p+ip$ BCS wavefunction\emph{ }\citep{Greiter91,read-green00}. 

\begin{figure}
\begin{longtable}{c}
\includegraphics[clip,width=0.57\columnwidth]{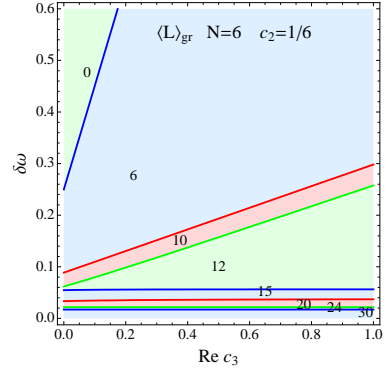}\tabularnewline
\end{longtable}

\caption{GS phase diagram for $N=6$ particles and $c_{2}=1/6$. Different
colors correspond to different angular momenta of the global GS. We
identify the Laughlin state ($\nu=1/2$) with $L=N(N-1)=30$, the
Pfaffian ($\nu=1$) with $L=N(N-2)/2=12$,  the single vortex with
$L=N$ and finally $L=0$. \label{fig:GroundRainbow}}

\end{figure}

The interplay between 2- and 3-body terms and the imperfect matching
$\delta\omega>0$ of trapping and rotation frequencies enriches the
phase diagram, by stabilizing other states aside (\ref{eq:laughlin}-\ref{eq:moore-read})
\citep{wilkin00}. The sequence of GSs has been computed numerically
by exact diagonalization in the case of elastic 3-body repulsion;
the result is presented in Fig.\ref{fig:GroundRainbow} for $N=6$,
in a truncated LLL single particle basis that contains all the angular
momenta up to $l_{\mathrm{max}}=2(N-1)$. This is enough to describe
exacly $\Psi_{\mathrm{Lau}}$, hence suitable also for the description
of states with lower angular momenta. In the LLL, central contact
potentials (\ref{eq:2-body delta potential}-\ref{eq:3-body delta potential})
can be expressed in terms of a single pseudopotential acting only
on zero relative angular momenta \citep{Haldane83}. Thus their sum
can be recast in a very compact form \citep{rizzi_future} \begin{eqnarray}
\mathcal{H}_{\mathrm{int}} & =\mathcal{H}_{2}+\mathcal{H}_{3}= & \sum_{n=2,3}\sum_{l}\gamma_{nl}d_{nl}^{\dagger}d_{nl}\label{eq:Hint}\end{eqnarray}
where $d_{nl}$ is an annihilation operator of $n$ particles with
total angular momentum projection $l$. The GS energy $E_{0}(L)$
is a monotone nonincreasing function of $L$. The global GS simply
belongs to the sector of $L$ that minimizes the quantity $E_{0}(L)+L\,\delta\omega$.
Notice that the separation lines in Fig. \ref{fig:GroundRainbow}
are nearly straight. In absence of the 3-body term, $c_{3}=0$, the
GS of $\mathcal{H}_{2}$ with $L=L_{\mathrm{Pf}}=12$ - that we indicate
as $|\psi_{0}^{(2)}(12)\rangle$ for brevity - is unique and stable
for a narrow interval of $\delta\omega.$ The state $|\psi_{0}^{(2)}(12)\rangle$
and the Pfaffian state (\ref{eq:moore-read}) share some similarities,
since they have the same angular momentum and their relative fidelity
is $F=|\langle\Psi_{\mathrm{Pf}}|\psi_{0}^{(2)}(12)\rangle|^{2}\simeq0.803$.
The most important feature of 3-body interaction is the enlargement
of both the stability interval and the fidelity, up to $F\simeq0.991$,
for $c_{3}=1$. 

The previous analysis urges us to design a mechanism for independent
tuning of 2- and 3-body terms, a task we tackle with the following
dimensional argument. In accordance with the fact that $\delta^{(2)}$
has dimensions of a inverse squared length $[\ell^{-2}]$, the coupling
$g_{2}^{\mathrm{2D}}$ in (\ref{eq:2-body delta potential}) has the
dimensions of an energy times a squared length. So, under rescaling
$E\rightarrow E/\hbar\omega$, $\ell\rightarrow\ell/\xi$ the adimensional
coupling $c_{2}=\sqrt{8\pi}a/\xi_{\perp}$ does not depend anymore
on the planar trap frequency. In other words, for 2-body collisions
2D are special since density does not discriminate between weakly
and strongly interacting regime. Along the same line, the adimensional
3-body coupling scales as the inverse of the effective trap area,
i.e. $c_{3}=g_{3}^{\mathrm{2D}}\xi^{-4}/\hbar\omega=(g_{3}^{\mathrm{2D}}m/\hbar^{2})\xi^{-2}.$
This suggests that the relative importance of 3-body collisions can
be boosted by a squeeze of the 2D trap, which increases the density.

Unfortunately, elastic collisions involving 3-body processes are rather
rare in nature. The most prominent 3-body collision process known
in physics of bosonic condensates is due to recombination \citep{Kraemer06}.
The formation of a biatomic molecule is assisted by a third particle
that assures energy-momentum conservation. The recombination rate
displays a rich behavior as a function of the 2-body scattering length
$a$, typically tuned via Feshbach resonances. For $a>0$ it shows
a universal behaviour $\sim a^{4}$ \citep{Esry99}, whereas for $a<0$
some genuine 3-body resonances are appearing due to Efimov trimer
states \citep{Kraemer06}. Interestingly, 3-body processes are still
present even in absence of 2-body scattering ($a=0$). This nonlinear
features allows to tune $c_{3}/c_{2}$, in conjuction with the squeezing
mechanism exposed above. Tipically, recombination is considered an
unwanted effect in experiments with condensates since it yields to
severe 3-body losses. Nonetheless, strong dissipation has been exploited
successfully to induce strong 2-body correlations in the 1D Tonks-Girardeau
gas, as observed in recent experiments \citep{syassen_science08,Garcia-Ripoll09}.
Moreover, 3-body dissipation has been proposed for obtaining a dimer
superfluid phase in 1D attractive bose-Hubbard models \citep{Daley09}.
On the same line, we propose here to use strong 3-body recombination
rate to filter out the Pfaffian state (\ref{eq:moore-read}), with
high fidelity and paying the small price of moderate losses.

The Markovian dynamics of the system is described by a Lindblad master
equation for the density matrix $\rho$\begin{equation}
\dot{\rho}=-\frac{i}{\hbar}\left[\mathcal{H}_{\mathrm{eff}}\rho-\rho\mathcal{H}_{\mathrm{eff}}^{\dagger}\right]+\sum_{l}\gamma_{3l}d_{3l}\rho d_{3l}^{\dagger},\label{eq:Lindblad}\end{equation}
where $\gamma_{3l}$ is the rate of decay in the channel of 3-body
total angular momentum $l$. The effective non-hermitian\emph{ }Hamiltonian
is given by \begin{equation}
\mathcal{H}_{\mathrm{eff}}=\mathcal{H}_{\mathrm{trap}}+\mathcal{H}_{2}-\frac{i}{2}\sum_{l}\gamma_{3l}d_{3l}^{\dagger}d_{3l}=\mathcal{H}_{\mathrm{trap}}+\mathcal{H}_{2}-\frac{i}{2}\mathcal{H}_{3}.\label{eq:Heff}\end{equation}
In the\textbf{ }quantum jump approach \citep{dalibard92,Dum92}, loss
events are assumed to be rare and are ideally detected by performing
very frequent measurements. This allows to describe the non-unitary
evolution of the system in terms of wave functions, through the equation
$i\hbar\left|\dot{\psi}(t)\right\rangle =\mathcal{H}_{\mathrm{eff}}\left|\psi(t)\right\rangle $.
The reduction of $\left\Vert \psi(t)\right\Vert ^{2}$ gives the probability
of having suffered from a jump, i.e. a 3-body loss, in $\left[0,t\right]$.
Actually, we are interested only in lossless samples and we discard
the cases where losses occur, since generally they lead to projection
onto excited states. Experimentally, the discarding procedure is made
possible by using post-selection of the samples, after measuring the
number of particles. In the limit case $c_{2}=0$, the dynamics is
governed only by the dissipative term and $\left|\psi(t)\right\rangle =\exp\left(-t\mathcal{H}_{3}\right)\left|\psi(0)\right\rangle $.
Due to the positiveness of $\mathcal{H}_{3}$, this evolution projects
onto K3 for long enough times, whose scale is set by $1/c_{3}$. 

Let us assume now to have prepared several copies of a few body system
with predominant 2-body collisions in the GS at angular momentum $L_{\mathrm{Pf}}$.
We will discuss later the experimental feasibility of this assumption.
At $t=0$, strong 3-body dissipation is suddenly switched on by squeezing
the trap. The ideal situation is accomplished by turning off completely
$c_{2}$ while having $c_{3}$ not too small. In this case the filtering
produces the Pfaffian state since it is unique in K3 for $L_{\mathrm{Pf}}$.
This works provided the starting $|\psi_{0}^{(2)}(L_{\mathrm{Pf}})\rangle$
has a sizable overlap with the Pfaffian, which indeed occurs for a
moderate $N$.

\begin{figure}
\begin{tabular}{>{\centering}p{0.5\columnwidth}>{\centering}p{0.5\columnwidth}}
\includegraphics[width=0.5\columnwidth]{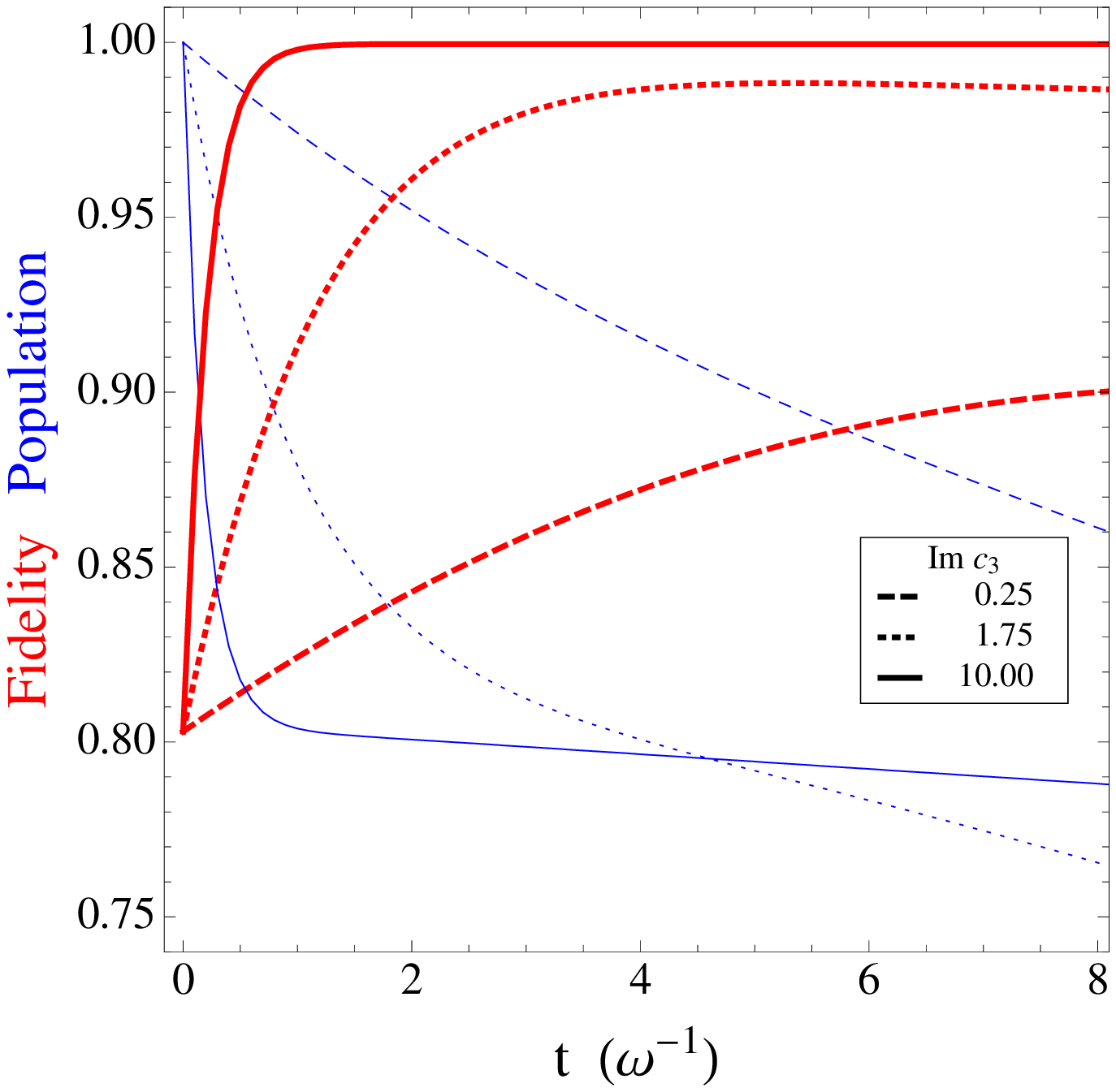} & \includegraphics[width=0.47\columnwidth]{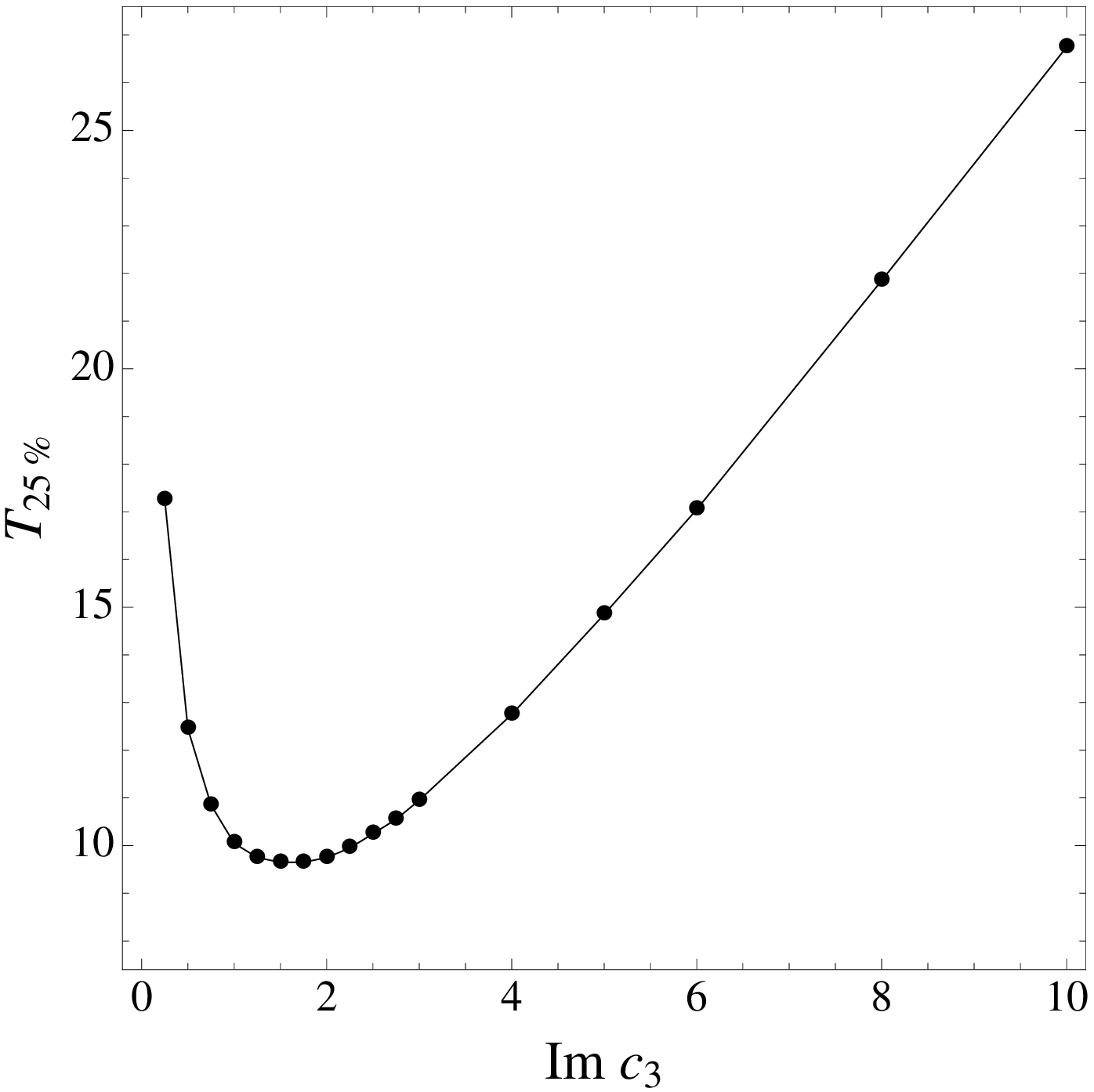}\tabularnewline
\end{tabular}

\caption{Left panel: fidelity with the Pfaffian and surviving population as
a function of time, after switching on dissipation $c_{3}$ on the
initial state $|\psi_{0}^{(2)}(12)\rangle$. The fidelity reaches
0.9994 for $c_{3}=10$. \label{fig:pop-fidel time} Right panel: threshold
time for having lost 25\% of population after switching the dissipation
$c_{3}$. \label{fig:Tpop25 c3}}

\end{figure}

Possibly, a more feasible experimental procedure would avoid to switch
off $c_{2}$ at $t=0$. The presence of 2-body collisions contrasts
the formation of pairing contained in the Pfaffian. However, we expect
that for $c_{3}/c_{2}\gg1$ the perfect projection is almost recovered.
In order to check this statement, we have numerically simulated the
evolution for $N=6$ by using a fourth-order Runge-Kutta algorithm.
In the left panel of Fig.\ref{fig:pop-fidel time} the population
of trajectories unaffected by jumps and the fidelity with the exact
Pfaffian state (\ref{eq:moore-read}) are plotted as a function of
time for different values of $c_{3}$. It emerges the following scenario:
strong enough dissipative rate yields to intense losses up to a time
after which the population is substantially stationary. The stabilization
of losses is a signal that the filtering procedure has converged to
a state very close to the Pfaffian one, as witnessed by the saturation
of fidelity to a value close to 1. This behaviour may be interpreted
as a sort of Quantum Zeno effect, where strong dissipation freezes
the system in K3, suppressing losses \citep{syassen_science08}. The
typical time we have to wait for reaching the threshold of $25\%$
losses is shown in the right panel of Fig.\ref{fig:Tpop25 c3}. As
expected, losses are not completely suppressed for long times, since
the Pfaffian is not an eigenstate of $\mathcal{H}_{2}$. 

The considerable improvement in the reproduction of the Pfaffian state,
with infidelity sinking from 20\% to almost 0.05\%, though relevant
by itself, does not exhaust all the importance of the proposed scheme.
Indeed the flurry of interest about this state is mainly related to
its zero-energy excitations, that have non-Abelian braiding properties
and may constitute the basic ingredients for TQC \citep{DasSarmaRevModPhys08}.
We now show that our filtering procedure allows for the production
and manipulation of the typical {}``half-flux'' quasiholes\begin{equation}
\Psi_{\mathrm{Pf+2qholes}}=\mathrm{Pf}\left(\frac{(z_{i}-w_{1})(z_{j}-w_{2})+(i\leftrightarrow j)}{z_{i}-z_{j}}\right)\prod_{i<j}(z_{i}-z_{j})\,,\label{eq:2qholes}\end{equation}
otherwise inaccessible with only rotation and conservative 2-body
interactions. The quasihole identification and motion is indeed possible
only in presence of an appropriate gap protected subspace, i.e. states
in K3 with $L>L_{\mathrm{Pf}}$. Such states are obviously steady
states of the 3-body dissipation, and the quasi-constant gap guarantees
that the filtering have almost the same speed and neatness for all
of them. Then, starting from having filtered the Pfaffian out of an
initial state, it would be in principle possible to engineer some
excitation scheme being sure to lie inside K3. However, due to the
80\% similarity between the GS  $|\psi_{0}^{(2)}(L_{\mathrm{Pf}})\rangle$
of $\mathcal{H}_{2}$ and the Pfaffian $|\psi_{\mathrm{Pf}}\rangle$,
one might be tempted to speculate an approximate scheme for quasiholes
with a similar precision. This is not the case, since the degeneracy
of the quasihole subpace is completely spoiled out, to the point we
cannot speak about a manifold protected by a gap. For testing our
assertion, we choose the state (\ref{eq:2qholes}) with quasiholes
located at the center, i.e. $w_{1}=w_{2}=0$, living in the sector
$L=L_{\mathrm{Pf}}+N$. By expanding (\ref{eq:2qholes}) in terms
of the eigenvectors of $\mathcal{H}_{2}$, we discover that this peculiar
state has a sizable overlap with several excited states of $\mathcal{H}_{2}$,
whose energies are spread over an interval of the order of $c_{2}$.

For completeness, we want to discuss about the experimental detection
of the Pfaffian state, that naturally follows the preparation. Of
course, a great evidence is provided by the suppression of losses,
that signals the absence of local 3-body superpositions. The probe
is conclusive after a measurement of $L$ and $N$, since the unicity
of the Pfaffian in K3. The Pfaffian is also expected to have an increased
pairing with respect to $|\psi_{0}^{(2)}(12)\rangle$. As a matter
of fact, the expectation value of the 2-body contact potential $\langle\mathcal{H}_{2}\rangle$
is respectively $3.13\: c_{2}$ and $3.31\: c_{2}$, with a difference
of about 5\%, maybe not striking enough to be resolved in a clean
way in experiments.

\begin{figure}
\begin{longtable}{c}
\includegraphics[clip,width=0.75\columnwidth]{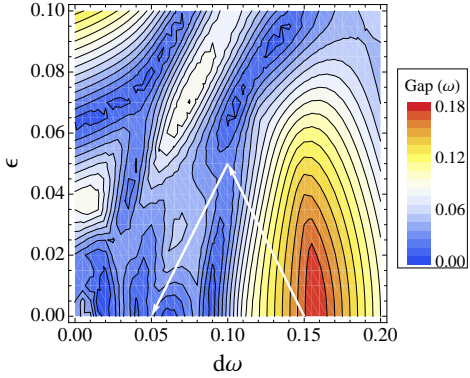}\tabularnewline
\end{longtable}

\caption{Map of the gap between GS and first excited state as a function of
the trap parameters $(\delta\omega,\epsilon)$. The white arrow indicate
the chosen path for conneting adiabatically the GS with $L=6$ and
the GS with $L=12$. The the matrix elements $|\langle\psi_{0}|\dot{H}|\psi_{m}\rangle|$
(as needed by the adiabatic criterion) give a very similar picture.\label{fig:gapmap}}

\end{figure}

In the previous study, we assume that before applying the filtering
via dissipation, it is possible to initialize the system in $|\psi_{0}^{(2)}(L_{\mathrm{Pf}})\rangle$.
To obtain a given filling factor $\nu$ it is necessary to apply a
rotation to the ultracold gas. The widely used experimental technique
consists on stirring the condensate \citep{schweikhard04,bretin04}.
In the rotating frame, it is equivalent to induce a small quadrupole
deformation to the trap $H\to H+H_{\epsilon}$, with\begin{equation}
H_{\epsilon}=\epsilon(x^{2}-y^{2}),\label{eq:Hepsilon}\end{equation}
where $\epsilon$ is meant to be small for avoiding heating and coupling
with higher LLs. The term (\ref{eq:Hepsilon}) couples single particle
states with angular momenta differing by 2. A possible route now is
to find an adiabatic path in the plane $(\delta\omega,\epsilon)$
connecting an already realized state, such as the single vortex GS
$|\psi_{0}^{(2)}(N)\rangle$, to the GS in the desired sector $|\psi_{0}^{(2)}(L_{\mathrm{Pf}})\rangle$.
Once again we resort to the case of $N=6$ and diagonalize the problem
in the full LLL Hilbert space in order to draw the map of the first
excitation gap \citep{popp04}, as displayed in Fig.\ref{fig:gapmap}.
A good path would try to avoid regions with a small gap, since they
are roughly associated to a slow down of the adiabatic evolution.
We have studied numerically the time evolution along the path marked
by the white arrows in Fig.\ref{fig:gapmap}, with an adaptive method
that adjusts the parameter speeds according to the adiabatic condition
$|\langle\psi_{0}|\dot{H}|\psi_{m}\rangle|\ll|E_{m}-E_{0}|^{2},\forall m\neq0$.
At the end of the path, it is possible to reach a fidelity of 99.7\%
with the GS at $L=12$. The price to pay is a evolution time $t\approx10^{3}\omega^{-1}$
which is quite long, but still comparable with the typical time of
such experiments in traps. The reason is that for reaching high angular
momenta, it seems unavoidable to cross regions with a small gap. The
way how Fig.\ref{fig:gapmap} scales with $N$ is not clear, and it
may be the case that the mean-field and QHE regimes are separated
by a quantum phase transition. This fact could explain why so far
experiments failed in reaching the QHE regime \citep{bretin04,schweikhard04,popp04}.
However, for a small number of particles like in our case, the gap
is still sizable and we expect experiments with the single holes of
optical lattices \citep{popp04} to succeed in the next future. Any
other scheme of GS preparation would be equally good to provide the
starting point for our procedure.

It is remarkable that the filtering strategy is working best if $c_{3}$
is turned on \emph{after} the preparation of $|\psi_{0}^{(2)}(L_{\mathrm{Pf}})\rangle$.
Our simulations indicate that switching on $c_{3}$ slowly or along
the path for preparing $|\psi_{0}^{(2)}(L_{\mathrm{Pf}})\rangle$
produce more losses. Finally, we emphasize that starting the adiabatic
preparation from states with higher $L$, like $\Psi_{\mathrm{Lau}}$,
needs evolution times which are too long, due to crossing of regions
with very small gaps.

In conclusion, we propose a feasible filtering scheme for the realization
and stabilization of Pfaffian state and excitations. Once the present
considerable efforts about preparing a system in the desired angular
momentum would achieve their goal, any other requirement is far within
the present technologies. We showed that a tuning mechanism for increasing
the relative importance of 3-body losses relies indeed on squeezing
the harmonic trap. Experiments in this direction have been done recently
\citep{Schneider_science08}. Reducing the magnitude of 2-body collisions,
by Feshbach resonance techniques, further enhances the filtering procedure.
The almost perfect cancellation of $c_{2}$ have been obtained in
Ref.\citep{roati_nature08}\emph{. }Detailed calculations and discussions
are included in a forthcoming work \citep{rizzi_future}. 

\begin{acknowledgments}
\emph{Acknowledgments} - We thank S. D\"urr, F. Minardi and G. Rempe
for discussions. We acknowledge support from the EU-IP SCALA %
\begin{comment}
under contract no. 015714
\end{comment}
{} and the DFG Excellenzcluster NIM. 
\end{acknowledgments}

\bibliographystyle{apsrev}
\bibliography{hall}

\end{document}